\documentclass[12pt,aps,prb,floats,eqsecnum,amsmath]
{revtex4}
\usepackage[dvips]{graphicx}
\renewcommand{\min}{\mathop{\rm min}\nolimits}

\def\lapprox{\,\raise0.4ex\hbox{$<$}\kern-0.8em\lower0.7ex\hbox{$\sim$}\,}
\def\gapprox{\,\raise0.4ex\hbox{$>$}\kern-0.8em\lower0.7ex\hbox{$\sim$}\,}

\begin{document}
\bibliographystyle{prsty}
\centerline{\large\bf Bound States in a Quantized Hall Ferromagnet}
\vskip 3mm { \centerline{S. Dickmann} }

\centerline{\it Institute for Solid State Physics of RAS, Chernogolovka, 142432 Moscow District, Russia}

\vskip 10mm

{}~{}~{}~{}~{}~{}\parbox{14.2cm} {\rm\small

We report on a study of the quasielectron-quasihole and
skyrmion-antiskyrmion bound states in the $\nu\!=\!1$ quantum Hall
regime. The short range attraction potential is assumed to be
determined by a point magnetic impurity. The calculations are
performed within the strong field approximation when the binding
energy and the characteristic electron-electron interaction energy
are smaller than the Landau level spacing.  The Excitonic
Representation technique is used in that case.}

\vskip 10mm

{\bf I.} Unique properties of the two-dimensional electron gas
(2DEG) in strong magnetic fields attract much attention to its
spectrum. In particular, the interaction of 2DEG with neutral
short range impurities exhibits unusual features in comparison
with its 3D prototype. \cite{an82,av93} In this paper we study the
bound fermion states appearing in the {\it quantum Hall
``ferromagnet''} (QHF) regime; i.e. the filling factor is
$\nu={\cal N}/N_{\phi}\simeq 2n\!+\!1$, where ${\cal N}$ and
$N_{\phi}=L^2/2\pi l_B^2$ are the numbers of electrons and
magnetic flux quanta ($L^2$ is the 2DEG area, $l_B$ is the
magnetic length). In the high magnetic field limit, which really
represents the solution to the first order in the ratio
$r_c=(e^2/\varepsilon l_B)/\hbar\omega_c$ considered to be small
($\omega_c$ is the cyclotron frequency, $\varepsilon$ is the
dielectric constant), we get the ground state with zeroth, first,
second,... and $(\!n\!\!-\!\!1\!)$-th Landau levels (LLs) fully
occupied and with $n$th level filled only by spin-up electrons
aligned along ${\bf B}$.

In the clean limit, fermion excitations are classified by their
spin-numbers $|\Delta S_z|\!=\!K\!+\!1/2$ ($K$ is
an integer),\cite{pa96} ranging from the simplest $|\Delta
S_z|\!=\!1/2$ case of quasielectrons or quasiholes to the
$K\!\to\!\infty$ limit which corresponds to the so-called
skyrmions. Certainly, the total energy of excitations incorporates
the Zeeman energy $|\epsilon_Z\Delta S_z|$, and the spin number of
lowest-lying fermions is thus determined by the actual value of
the gap $\epsilon_Z\!=\!|g\mu_BB|$. As regards to the Coulomb
exchange energy of the fermions, in the $\nu\!=\!1$ case this part
of the total energy decreases monotonically with the $K$
number.{}\cite{pa96} E.g., to the first order in
$r_{\mbox{\tiny C}}$ the exchange energies are
$-\frac{1}{4}\sqrt{\pi/2}e^2/\varepsilon {}l_B$ and
$\frac{3}{4}\sqrt{\pi/2}e^2/\varepsilon {}l_B$ (in the strict 2D
limit) for electron-like and hole-like skyrmions, respectively.
\cite{so93,fe94,by96,io99,di02} For comparison, in the $K\!=\!0$
case these ones are $0$ (electrons) and
$\sqrt{\pi/2}e^2/\varepsilon {}l_B$ (holes).

We calculate the bound-state energies in the presence of point
impurities.  In the single electron approximation (i.e. in the
$\nu\!\ll 1$ case) this problem was studied in Ref.
\onlinecite{av93} for arbitrary strengh of the perpendcular
magnetic field. In Refs. \onlinecite{fl01} and \onlinecite{da02}
the authors investigated the genesis of the impurity potential in
the two-dimensional channel and gave a proof for the short range
approach. The latter in terms of the ``envelope'' wave-function
method means that the inpurity Hamiltonian can be modeled by the
$\delta$-function:
$$
  {\hat H}_{\mbox{\scriptsize imp}}=2\pi\left(W{\hat{\bf 1}}+\frac{D}{2}\hat{\sigma_Z}\right)
  \delta({\bf r})        \eqno (1)
$$
(here and in the following $\sigma_{x,y,z}$ are the Pauli matrices).
The $D\not= 0$ case corresponds to a paramagnetic impurity with its own
magnetic moment aligned parallel to the magnetic field. (We consider that
${\vec B}\parallel {\hat z}$.) At the negative $g$-factor the positive value
of $D$ provides capture of the spin wave (spin exciton) by an impurity in
the QHF case.\cite{fl01}

We will solve the problem in the ``shallow'' impurity approximation,
$$
  |W_{\sigma}|\ll \hbar^2/m_e^*\qquad (\sigma\!=\!\uparrow,\downarrow),   \eqno(2)
$$
where $W_{\uparrow,\downarrow}\!=\!W\!\pm\!D/2$. Actually one will
see that this condition enables to employ the high magnetic field
approach:
$$
  E_b\ll \hbar\omega_c\,,        \eqno(3)
$$
where $E_b$ is the desired binding energy, and $\omega_c$ is the
cyclotron frequency. The condition (3) allows us to employ the
projection onto a single LL approach when calculating $E_b$ in the
leading approximation. At the same time we still consider that the
point-impurity approximation is not disturbed, i.e. the  impurity
localization radius $\rho_b$ is much smaller than the magnetic
length: $\rho_b\ll l_B$.

{\bf II.} Any single-electron state may be presented in the form of the
expansion
$$
  {\chi}=\sum_{a p}c_{a p}{ \phi}_{a p},   \eqno (4)
$$
where we choose the Landau-gauge functions $\phi_{ap}$ as the
basis set. The subscript $p$ distinguishes between different
states belonging to a continuously degenerate Landau level, and
the label $a$ is a binary index $a=\left(n_a,\sigma_a\right)$,
which represents both LL index and spin index. We have thus
${\phi}_{ap}({\bf
r},\sigma)=\delta_{\sigma,\sigma_a}(l_BL)^{-1/2}e^{ipy}
  \varphi_{n_a}(pl_B+x/l_B)$, where
$$
  \varphi_{n}(x)=
  (2^{n}n!\sqrt{\pi})^{-1/2}e^{-x^2/2}\mbox{\sl H}_{n}(x)  \eqno (5)
$$
[$\mbox{\sl H}_{n}(x)$ is the Hermite polynomial]. In the
following we employ the notation ${a}_p,\;\,{b}_p$,... for the
electron annihilation operator corresponding to sublevel
$a,\;\,b$,... and use also the intra-LL ``displacement'' operators
${\cal A}_{\bf q}^+,\;\,{\cal B}_{\bf q}^+$,..., where
$$
  {{\cal A}}_{{\bf q}}^{+}={N_{\phi}}^{-1}\sum_{p}\,
  e^{-iq_x pl_B^2}
  a_{p+\frac{q_y}{2}}^{+}\,a_{p-\frac{q_y}{2}},\qquad
  {\cal B}_{\bf q}^{+}=(a\to b),...     \eqno (6)
$$
(${\cal A}_{{\bf q}}^{+}={\cal A}_{-{\bf q}}$). Considering the
quantity $\chi$ as an annihilation   operator in the Schr\"odinger
representation, we can substitute Eq. (4) into $\langle\chi|{\hat
H}_{\mbox{\scriptsize imp}}|\chi\rangle$ to obtain the secondary
quantized representation of the Hamiltonian ${\cal
H}_{\mbox{\scriptsize imp}}$, namely:
$$
  {\hat {\cal H}}_{\mbox{\scriptsize imp}}\approx l_B^{-2}\sum_{\bf q}e^{-q^2l_B^2/4}
  \left(W_{\uparrow}{\cal
  A}^+_{\bf q}+W_{\downarrow}{\cal B}^+_{\bf q}\right).                 \eqno (7)
$$
Here the sign of approximate equality means that we have omitted
the terms corresponding to the LL mixing and have kept only those
relevant to the projection onto the $n$th LL. Therefore,
specifically we have in Eq. (7) that the labels $a$ and $b$
correspond to $a=(n,\uparrow)$ and $b=(n,\downarrow)$ sublevels.
At $\nu\!=\!2n\!+\!1$ the ``clean" ground state $|0\rangle$ is
completely determined by the equations ${\cal A}_{{\bf
q}}|0\rangle\!= \!\delta_{{\bf q},0}|0\rangle$ and ${\cal B}_{{\bf
q}}|0\rangle\!=\!0$, while electron and hole are defined as the
$$
  |f_e\rangle=\sum_p f_e(p)b_p^+|0\rangle\quad \mbox{and} \quad
  |f_h\rangle=\sum_p f_h(p)a_p|0\rangle                \eqno(8)
$$
states, respectively. Here the envelope functions are normalized as
$\sum_p|f_{e,h}|^2=1$.

The quasiparticle states (8) satisfy the ``clean'' equations
${\hat{\cal H}}_0|f_{e,h}\rangle= (E_0+E_{e,h})|f_{e,h}\rangle$,
where $E_e=\epsilon_Z$ and $E_h\!=\!E_C$. Here $E_C$ is the
characteristic Coulomb energy: $E_C\!=\!\int e^{-q^2l_B^2/2}V({\bf
q})d{\bf q}$, where $2\pi V({\bf q})$ is the 2D Fourier component
of the averaged Coulomb potential (in the ideal 2D case
$V\!=\!e^2/\kappa q$ and $E_C\!=\!\sqrt{\pi/2}e^2/\kappa l_B$). To
obtain this result, it is convenient to employ the expression for
the Coulomb interaction Hamiltonian in terms of the Excitonic
Representation (see, e.g., Ref. \onlinecite{di02}) and the
commutation rules
$$
  \left[{\cal A}^+_{\bf q},a_p\right]\equiv-\frac{1}{N_{\phi}}
  e^{-iq_xl_B^2(p\!-\!q_y/2)}
  a_{p\!-\!q_y},\quad \left[{\cal B}^+_{\bf q},b_p^+\right]\equiv
  \frac{1}{N_{\phi}}e^{-iq_xl_B^2(p\!+\!q_y/2)}b_{p\!-\!q_y}^+
  \eqno (9)
$$
and $\left[{\cal A}^+_{\bf q},b_p^+\right]=
\left[{\cal B}^+_{\bf q},a_p\right]\equiv 0$.

First we obtain the correction to the ground state energy in the
case of single impurity. Substituting expression (7) for
${\hat{\cal H}}_{\mbox{\scriptsize imp}}$ into the equation $
  \left({\hat{\cal H}}_0+
  {\hat{\cal H}}_{\mbox{\scriptsize imp}}\right)|0\rangle=
  (E_0+\Delta E_0)|0\rangle
$ (where ${\hat{\cal H}}_0$ is the ``clean" QHF Hamiltonian {\it
including the Zeeman and Coulomb interaction part}) we obtain the
correction to the ``clean" ground state: $\Delta
E_0=W_{\uparrow}/l_B^2$.

Finally, to calculate the bound electron state we solve the
equation ${\hat{\cal H}}_{\mbox{\scriptsize
imp}}|f_e\rangle=E'_e|f_e\rangle$, which with help of Eqs. (7)-(9)
is reduced to the integral equation
$$
  \frac{W_{\downarrow}}{l_B\sqrt{\pi}}\int_{-\infty}^{+\infty}dsf_e(s)
  e^{-(s^2\!+\!p^2)l_B^2/2}=E'_ef(p).                    \eqno (10)
$$
The latter has the solution $f_e(p)=(\pi N_{\phi})^{-1/4}e^{-p^2l_B^2/2}$ at
$E'_e=W_{\downarrow}/l_B^2$
[$(\pi N_{\phi})^{-1/4}$ is the normalization factor].
Hence the electron binding energy is
$$
  E^{(e)}_b=-W_{\downarrow}/l_B^2.    \eqno (11)
$$
Naturally, this state is realized if $W_{\downarrow}< 0$. Let us
note that in the leading approximation the obtained $E^{(e)}_b$
value is equal to the binding energy in the single electron
problem.\cite{av93} In a similar way we obtain that the equation
${\hat{\cal H}}_{\mbox{\scriptsize
imp}}|f_h\rangle=E'_h|f_h\rangle$ has the solution $f_h(p)=(\pi
N_{\phi})^{-1/4}e^{-p^2l_B^2/2}$ at $E'_h=-W_{\uparrow}/l_B^2$.
So, the total energy of the $|f_h\rangle$ bound state is
$E_0\!+\!E_C$; i.e. it is just the same as in the ``clean'' hole
state. The bound energy of the hole is
$$
  E^{(h)}_b=W_{\uparrow}/l_B^2.     \eqno (12)
$$
This state exists under the condition $W_{\uparrow}> 0$.

The physical meaning of the envelope functions $f_{e,h}$ obtained
above becomes evident if we change the Landau gauge to the
symmetric gauge. In the latter case we have to change the vector potential
$\mbox{\boldmath $A$}$= $(0,Bx,0)$ for $\mbox{\boldmath
$A$}$=$(-By/2,Bx/2,0)$. Then the single electron states of the $n$'s
LL are described by the basis spatial function
$$
  \phi_{nm}=l_B^{-1}\left[\frac{n!}{2^{m+1}(m+n)!\pi}\right]^{1/2}(ir/l_B)^m
  L^m_{n}(r^2/l_B^22)e^{-im\varphi-r^2/4l_B^2}\quad (n+m\geq 0), \eqno (13)
$$
where $\mbox{\boldmath $r$} =(r\cos{\varphi},r\sin{\varphi})$,
$L_n^m$ is the Laguerre polynomial and $m$ runs over ${ N}_{\phi}$
integer numbers: $m=-n,\,1\!-\!n,\,
2\!-\!n,...,\,N_{\phi}\!-\!n\!-\!1$. All these states have the
same cyclotron energy, and the Fermi creation and annihilation
operators acquire index $m$ (instead of $p$ in the
Landau gauge) now. For example, one can find the expression for
old Landau gauge operator $a_p$ in terms of new operators $a_m$:
$$
  a_p={N_{\phi}}^{-1/2}\sum_{m=0}^{{\cal N}-1}i^{m\!-\!n}
 \varphi_m(pl_B)a_{m-n}  \eqno (14)
$$
where $\varphi_m$ is the oscillatory function (5). Now, if we
substitute this expression (and analogous one for $b_p^+$) into
Eqs. (8) with the above found functions, we obtain that only the $m=0$
harmonic of Eq. (13) contributes to the $|f_{e,h}\rangle$ bound
states. Indeed, the summation over $p$ in the expansion (8) turns
out to be proportional to the integral $\int_{-\infty}^{\infty}
e^{-p^2}H_m(p)$. The latter vanishes at any $m$ except for $m=0$.
Certainly, this feature reflects the well known fact that the
point impurity Hamiltonian is diagonal exactly in the symmetric
basis (13). Besides, only the zero (axially symmetric) harmonic
contributes to the bound state energy calculated within the single
LL approximation.

{\bf III.} To study a bound skyrmionic excitation we present it in accordance
with Refs. \onlinecite{io99,di02} as a smooth rotation in the three-dimensional
spin space
$$
  {\vec \psi}({\bf r})={\hat U}({\bf r}){\vec \chi}({\bf r}),\quad
 {\bf r}=(x,y)\,.           \eqno (15)
$$
Here ${\vec \psi}$ is a spinor given in the stationary coordinate
system and ${\vec \chi}$ is a new spinor in the local coordinate
system following this rotation. The rotation matrix ${\hat U}({\bf
r})\quad$ (${\hat U}^{\dag}{\hat U}=1$) is parameterized by three
Eulerian angles:\cite{ll91}
$$
  {\hat U}=\left(
  \begin{array}{cc}
  \cos{\frac{\theta}{2}}e^{-i(\varphi+\eta)/2}&\sin{\frac{\theta}{2}}e^{i(\eta-\varphi)/2}\\
  -\sin{\frac{\theta}{2}}e^{i(\varphi-\eta)/2}&\cos{\frac{\theta}{2}}e^{i(\varphi+\eta)/2}\\
  \end{array}
  \right).                             \eqno(16)
$$
These angles $\theta({\bf r})$, $\varphi({\bf r})$ and $\eta({\bf
r})$ present continuum field functions. The skyrmion state is thus
determined by the continuum matrix ${\hat U}({\bf r})$ and by the
local quantum state ${\vec \chi}$ determined in terms of small
gradient ($~l_B\nabla {\hat U}$) corrections to the local QHF,
where all electron spins are parallel [${\vec \chi}\propto
{1\choose 0}$].

The solution may be found from the reformulated variational
principle. Namely, we divide the 2DEG area into a large number
$G_i$ of domains which are much smaller than the total 2DEG area
but still remain much larger than the magnetic flux quantum area
$2\pi {}l_B^2$. The energy of excitations of this type may be
found through the minimization procedure in the following way:
$$
  E=\min\limits_{U}^{}\left[\sum_i\min\limits_{\psi}^{}
  \left(\frac{\left\langle\psi|H_i|\psi\right\rangle_{G_i}}
  {\left\langle\psi|\psi\right\rangle_{G_i}}\right)\right]. \eqno (17)
$$
Here averaging is performed over the domain $G_i$. All the $G_i$
areas add up to the total 2DEG area. $H_i$ is the Hamiltonian
corresponding to the $G_i$ domain. The state $|\psi\rangle$
presents here a many-electron quantum state built by the single
electron spinors (15). So, the state $|\psi\rangle$ is
parametrized by ${\hat U}$ and by the derivatives of ${\hat U}$
(generally, up to any order) considered as external parameters for
every $G_i$. The procedure of the inner minimization in Eq. (17)
is equivalent to the solution of the Schr\"odinger equation within
the area $\Delta x\Delta y=G_i$.

In our case the local Hamiltonians are just the same as in the
clean case except for the Hamiltonian $H_{i=0}$ corresponding to
the one domain $G_0$ which involves the point impurity. The
procedure of the minimization (17) only differs from the clean
case by adding the energy
$$
  \Delta E_{\mbox{\scriptsize imp}}^{(U)}=\frac{\langle\chi|{\hat U}^+
  {\hat H}_{\mbox{\scriptsize imp}}{\hat
  U}|\chi\rangle_{G_i}}{\langle\chi|\chi\rangle_{G_i}}    \eqno (18)
$$
to the sum within the square brackets in Eq. (17). Here we have
used Eq. (15) and then should substitute
$$
  {\hat U}^+{\hat H}_{\mbox{\scriptsize imp}}{\hat U}=
  \left(
  \begin{array}{cc}
  W+\frac{D}{2}\cos{\theta}&-\frac{D}{2}\sin{\theta}e^{-i\varphi}\\
  -\frac{D}{2}\sin{\theta}e^{i\varphi}&W-\frac{D}{2}\cos{\theta}\\
  \end{array}
  \right)\delta({\bf r})                     \eqno (19)
$$
into Eq. (18). (We have set $\eta\!=\!\varphi$ without any loss of
generality.) Under the conditions (2)-(3), it is sufficient to use
as $|\chi\rangle$ the unperturbed QHF state determined in the
local coordinate system of domain $G_0$. That is, in our case
$|\chi\rangle={\hat {\vec \chi}}|0\rangle$, where ${\hat {\vec
\chi}}$ is the annihilation operator. If employing again the
expansion of Eq. (4), then with the help of equation
$$
  {\hat{\cal  H}}_{\mbox{\scriptsize imp}}=\int
  \limits_{\mbox{\scriptsize over}\:
  G_0}d{\bf r}\,{\hat{\vec \chi}}^+ {\hat U}^+
  {\hat H}_{\mbox{\scriptsize imp}}
  {\hat U}{\hat{\vec \chi}}           \eqno (20)
$$
we can obtain the impurity Hamiltonian in terms of the secondary
quantized representation. We will study only the $\nu=1$ case.
Substituting Eqs. (19) and (4) into Eq. (20) and considering that
$a=(0,\uparrow)$ and $b=(0,\downarrow)$, we find within the single
LL approximation
$$
  \begin{array}{r}
  {\hat{\cal  H}}_{\mbox{\scriptsize imp}}=
  l_B^{-2}\sum_{\bf q}e^{-q^2l_B^2/4}\left[(W+\frac{D}{2}\cos{\theta_0}){\cal
  A}^+_{\bf q}+(W-\frac{D}{2}\cos{\theta_0}){\cal B}^+_{\bf q}\right.\\\left.-
  \frac{D}{2}\sin{\theta_0}e^{-i\varphi_0}N_{\phi}^{-1/2}{\cal Q}_{\bf q}
  -\frac{D}{2}\sin{\theta_0}e^{i\varphi_0}
  N_{\phi}^{-1/2}{\cal Q}^+_{\bf q}\right].\\
  \end{array}
  \eqno (21)
$$
Definitions of the operators ${\cal A}_{\bf q}^+$ and ${\cal
B}_{\bf q}^+$ are identical to those given by Eq. (6) with the
number of magnetic flux quanta $N_{\phi}={\Delta x\Delta y}/{2\pi
{}l_B^2}$ being non-zero only for the domain $G_0$. The notation
of the spin-exciton creation $ {{\cal Q}}_{{\bf
q}}^{+}={N_{\phi}}^{-1/2}\sum_{p}\,
  e^{-iq_x pl_B^2}
  b_{p+\frac{q_y}{2}}^{+}\,a_{p-\frac{q_y}{2}}$ and annihilation
${\cal Q}_{\bf q}=\left( {{\cal Q}}_{{\bf q}}^{+}\right)^+$
operators have also been used in Eq. (21). $\theta_0$ and
$\varphi_0$ are the Hermitian angles (determined by the given
${\hat U}$ matrix) corresponding to the domain $G_0$.
Now in accordance with Eq. (18) we get
$$
  \Delta E_{\mbox{\scriptsize imp}}^{(U)}=\langle 0|
  {\hat{\cal H}}_{\mbox{\scriptsize imp}}|0\rangle=
  \left(W+\frac{D}{2}\cos{\theta_0}\right)/l_B^2.        \eqno(22)
$$
(Only the $\sim {\cal A}_0^+$ term contributes to this result.)

The outer minimization in Eq. (17) is thereby presented as $
  \min\limits_{U}^{}\left(E_U+\Delta E_{\mbox{\scriptsize imp}}^{(U)}\right)
$, where $E_U$ is the ``clean'' energy obtained for a given
function ${\hat U}({\bf r})$ after the summation over all $G_i$ in
Eq. (17). Meanwhile the minimization of $E_U$ and $\Delta
E_{\mbox{\scriptsize imp}}^{(U)}$ may be fulfilled independently.
Indeed, the clean skyrmion state is degenerate having the energy
which does not depend on the ``skyrmion center'' position. The
latter is the point of the total 2DEG area where $\theta=\pi$,
i.e. local electron spins are aligned in the direction opposite to
the direction at the infinity (where $\theta=0$). In contrast to
this, the energy $\Delta E_{\mbox{\scriptsize imp}}^{(U)}$ just
depends on the relative positions of the impurity site (the {\bf
r}=0 point in our coordinate system) and the skyrmion center site.
Thus the impurity lifts the degeneracy. Hence by varying the
position of the skyrmion center site, we find that
$\min\limits_{U}^{}\left[\Delta E_{\mbox{\scriptsize
imp}}^{(U)}\right]= (W\!-\!|D|/2)/l_B^2$. If $D<0$, this value is
reached at $\theta_0=0$, i.e. the skyrmion center is located at
the infinity. Evidently in this case the skyrmion does not form a
bound state. At $D>0$, the minimum energy is realized for
$\theta_0=\pi$, i.e. where the impurity site and the skyrmion
center coincide (of course, to within the length smaller than the
characteristic skyrmion radius $R^*$ but perhaps larger than
$l_B$). This means that the bound skyrmion state takes place. The
binding energy should be obtained by subtraction of the minimum
energy from the $\theta_0=0$ energy $(W\!+\!D/2)l_B^2$, i.e.
$$
  E_b^{\mbox{\scriptsize {sk}}}=D/l_B^2.   \eqno(23)
$$
So, in the adopted approximation only the
magnetic impurity with $D>0$ captures the skyrmion (cf. Ref.
\onlinecite{fl01} where the similar result is obtained in the case
of the bound spin exciton). It is worth to note that the result
(23) does not depend on the skyrmion charge:\cite{foot}
electron-like and hole-like skyrmions have the same bound energy
(23).

Analysis reveals that the charge dependence arises only
in the second order approximation in terms of $l_B\nabla$,
being determined by the relative correction $~(l_B/R^*)^2$. This
can be easily found by means of the renormalization procedure for
the magnetic length $l_B\to {\tilde l_B}$. Indeed, the effective
local magnetic length is determined by the effective magnetic
field, including the additional part proportional to the second-order
spatial derivatives of the field ${\hat U}$:$\,$\cite{di02}
$$
  \frac{1}{{\tilde l}_B^2} =
  \frac{1}{l_B^2}+ {\vec \nabla}\times{\vec \Omega}^z\,.
  \eqno(24)
$$
Here
$$
  {\vec \Omega}^z=\frac{1}{2}\left(1+\cos{\theta}\right){\vec \nabla}
  \varphi   \eqno(25)
$$
(see Refs. \onlinecite{io99,di02}) and $l_B$ is the magnetic
length at the infinity (far from the skyrmion center). Let us
assume that the impurity is non-magnetic, i.e. $D=0$. The binding
energy (if the bound state would be realized) should be determined
exactly by the desired correction. We substitute into Eqs.
(24)-(25) for the angles their expressions in terms of functions
of ${\bf r}\,{}\,$\cite{be75}
$$
  \cos\theta=\frac{{R^*}^2-r^2}{{R^*}^2+r^2}\,,\quad\mbox{and}\quad
  \phi=-q\arctan(y/x)
$$
(where $q=\pm$ is the skyrmion charge). Eventually, by comparing
the skyrmion energy at ${\bf r}=0$ with the energy at the
infinity, we find the impurity correction: $\Delta
E_{\mbox{\scriptsize imp}}=W{\vec \nabla}\times
  \left.{\vec \Omega}^z\right|_{{\bf r}=0}=
  2qW/{R^*}^2$. The binding energy is thus
$$
  E_{b}^{\mbox{\scriptsize sk}}=-2qW/{R^*}^2
  \quad\mbox{if}\quad D=0\,.
$$
Therefore the bound electron/hole-like skyrmion arises when $W$
is positive/negative.

In reality, the concentration of point impurities in the 2D
channel can be considerable. At least it seems to have the values at
which the mean distance between impurities is well shorter than
the effective skyrmion radius (the latter is determined by small
but still non-zero Zeeman gap $\epsilon^*_Z$). In this case the
impurity contribution to the total energy is equal to $\int d{\bf
r}\lambda l_B^{-2}\left[W\!+\!\frac{D}{2}\cos{\theta({\bf r})}
\right]$, where $\lambda({\bf r})$ is the concentration of
impurities. It involves also the correction to the ground state energy
$l_B^{-2}\left(W\!+\!\frac{D}{2}\right)\int d{\bf r}\lambda$ which
should be subtracted. The impurity correction to the proper
skyrmion energy is thereby
$$
  \Delta E_{\mbox{\scriptsize imp}}^{\mbox{\scriptsize sk}}=\frac{D}{2l_B^2}\int
  \left[\cos{\theta({\bf r})}-1\right]\lambda d{\bf r}\,.
$$
If we compare this value with the skyrmion Zeeman energy
$$
  \frac{\epsilon_Z^*}{2}\int(1-\cos{\theta})d{\bf r}/2\pi l_B^2\,,
$$
it becomes evident that the magnetic impurities dispersed in the 2D
channel lead to the correction to the Zeeman gap:
$$
  \epsilon_Z^*\to \epsilon_Z^*-2\pi \lambda D
$$
(assuming a homogeneous concentration $\lambda$). Due to the supposed
smallness of $\epsilon_Z^*$, this can be substantial. Under certain
conditions it could change the sign of effective $g$-factor in the
2DEG.

I acknowledge support by the Russian Fund for Basic Research and
thank V. Fleurov, S.V. Iordanskii and K. Kikoin for useful
discussions.


\begin{references}
\bibitem{foot}
{The skyrmion electric charge is equal to the so-called
topological charge. We do not distinguish these two meanings here.}
\end{references}
\end{document}